\documentclass[11pt,a4paper]{article}

\usepackage[utf8]{inputenc}
\usepackage[english]{babel}
 
\usepackage{amsmath, mathtools}
\usepackage{stmaryrd}
\usepackage{multicol}
\usepackage[margin=1in]{geometry}
\usepackage{graphicx}
\usepackage{subcaption}
\usepackage{multirow}
\usepackage{xcolor,colortbl}
\usepackage{newtxtext}
\usepackage{newtxmath}
\usepackage{wrapfig}
\usepackage{tabularx}
\usepackage{hyperref}
\usepackage{authblk}

\usepackage{amsmath}
\DeclareMathOperator*{\argmax}{argmax}

\DeclareMathOperator*{\E}{\mathbb{E}}
\newcommand{\simpc}{S^{PC}}
\newcommand{\simed}{S^{ED}}
\newcommand{\simsp}{S^{SP}}
\newcommand{\simpcsp}{CS^{PC+SP}}
\newcommand{\simedsp}{CS^{ED+SP}}
\newcommand{\modelts}{\theta_{ts}}
\newcommand{\modelnet}{\theta_{net}}
\newcommand{\changej}{\bigtriangledown J}
\newcommand{\changeops}{\bigtriangledown OPS}
\usepackage{breqn}
\makeatletter
\patchcmd\eq@setnumber{\stepcounter}{\refstepcounter}{}{%
  \errmessage{Patching \noexpand\eq@setnumber failed}%
}

\newcommand\Autoref[1]{\@first@ref#1,@}
\def\@throw@dot#1.#2@{#1}
\def\@set@refname#1{
    \edef\@tmp{\getrefbykeydefault{#1}{anchor}{}}%
    \xdef\@tmp{\expandafter\@throw@dot\@tmp.@}%
    \ltx@IfUndefined{\@tmp autorefnameplural}%
         {\def\@refname{\@nameuse{\@tmp autorefname}s}}%
         {\def\@refname{\@nameuse{\@tmp autorefnameplural}}}%
}
\def\@first@ref#1,#2{%
  \ifx#2@\autoref{#1}\let\@nextref\@gobble
  \else%
    \@set@refname{#1}
    \@refname~\ref{#1}
    \let\@nextref\@next@ref
  \fi%
  \@nextref#2%
}
\def\@next@ref#1,#2{%
   \ifx#2@ and~\ref{#1}\let\@nextref\@gobble
   \else, \ref{#1}
   \fi%
   \@nextref#2%
}

\newcases{nullcases}
    {\quad}
    {\hfil{##}} {{##}\hfil}
    {.} {.}

\makeatother

\title{TiCoNE 2: A Composite Clustering Model for Robust Cluster Analyses on Noisy Data}
\author[1]{Christian Wiwie}
\author[1]{Richard Röttger}
\author[1,2]{Jan Baumbach}
\affil[1]{Institute of Mathematics and Computer Science, University of Southern Denmark, Campusvej 55, DK-5230 Odense, Denmark.}
\affil[2]{Chair of Experimental Bioinformatics, Technical University of Munich, Maximus-von-Imhof-Forum 3
85354 Freising-Weihenstephan, Germany.}

\begin{document}
\maketitle
\begin{abstract}
Identifying groups of similar objects using clustering approaches is one of the most frequently employed first steps in exploratory biomedical data analysis. Many clustering methods have been developed that pursue different strategies to identify the optimal clustering for a data set.

We previously published TiCoNE, an interactive clustering approach coupled with de-novo network enrichment of identified clusters. However, in this first version time-series and network analysis remained two separate steps in that only time-series data was clustered, and identified clusters mapped to and enriched within a network in a second separate step.

In this work, we present TiCoNE 2: An extension that can now seamlessly incorporate multiple data types within its composite clustering model. Systematic evaluation on 50 random data sets, as well as on 2,400 data sets containing enriched cluster structure and varying levels of noise, shows that our approach is able to successfully recover cluster patterns embedded in random data and that it is more robust towards noise than non-composite models using only one data type, when applied to two data types simultaneously. 

Herein, each data set was clustered using five different similarity functions into k=10/30 clusters, resulting to \textasciitilde5,000 clusterings in total. We evaluated the quality of each derived clustering with the Jaccard index and an internal validity score. We used TiCoNE to calculate empirical p-values for all generated clusters with different permutation functions, resulting in \textasciitilde80,000 cluster p-values. We show, that derived p-values can be used to reliably distinguish between foreground and background clusters.

TiCoNE 2 allows researchers to seamlessly analyze time-series data together with biological interaction networks in an intuitive way and thereby provides more robust results than single data type cluster analyses.

\end{abstract}
\begin{multicols}{2}

\section{Introduction}

Biomedical data sets collected in large-scale become increasingly complex and numerous. Biological systems are highly complex, and integrated analysis of multiple data sets together is needed to create high-quality models suited to gain novel insights and draw accurate conclusions.

A multitude of analysis algorithms have been developed, unsupervised clustering algorithms being some of the most widely used ones today \cite{BroheeEvaluationclusteringalgorithms2006,WittkopLargescaleclustering2007,SaltonDevelopmentsautomatictext1991,NavigliWordsensedisambiguation2009,Prado-Vazqueznovelapproachtriplenegative2019}. They can be employed in the initial step of any analysis pipeline, when no further knowledge is available to guide the analysis.

We previously published TiCoNE 1 \cite{doi:10.1089/sysm.2018.0013}, a human-guided, interactive clustering method for the conjoint analysis of time-series data with interaction networks. It employs a prototype-based, greedy clustering optimization scheme to identify overrepresented time-series patterns, and regions in a network that are enriched in these identified patterns by means of de-novo network enrichment with KeyPathwayMiner \cite{AlcarazRobustnovopathway2016}. Hence, the analysis of multiple data sets consisted of two separate steps that were performed after each other: (1) clustering time-series, and (2) de-novo network enrichment of the identified clusters.
We strongly believe, that incorporating the two data types more seamlessly already during the clustering procedure will lead to better and more accurate results.

In this work, we extend TiCoNE's core methodology such that its clustering procedures are now seamlessly integrating multiple data types, such as time-series data and biological interaction networks, without the need to perform de-novo network enrichment as a separate follow-up step.

We systematically evaluate TiCoNE's ability to recover cluster patterns embedded in random data and its robustness towards noise. To this end we generated a large set of time-series and network data sets with varying levels of noise, numbers of embedded clusters, and sizes of embedded clusters. Next, we investigated to which extend results from a conjoint analysis of two data types are superior over results from the analysis of a single data set.
Finally, we use TiCoNE's statistical facilities to calculate empirical p-values for all generated clusters using different available permutation functions. We systematically analyze whether TiCoNE's cluster p-values are a reliable means of attributing clusters with a level of importance, by ensuring that random clusters are assigned insignificant p-values, and planted foreground clusters can successfully be distinguished from background noise.
\section{Methods}
\subsection{Composite clustering model}

TiCoNE is a prototype based clustering approach, where each cluster $c_i$ of a clustering $C$ is associated with a prototype $p(c_i)$ that best represents its members $o_j \in c_i$.

While objects and cluster prototypes were simply modeled via a time-series in TiCoNE 1, we extended this model such that they are now composites and can consist of several components $\theta_i$, each corresponding to one data type.

For each clustering process we define a model $\theta=[\theta_1,\ldots,\theta_M]$, based on the used composite components. A clustering model can only be employed, if all component data types are available for the objects to be clustered.
We can then formally define the model of objects and prototypes in terms of their components as $\theta(o)=[\theta_1(o),\ldots,\theta_M(o)]$ and $\theta(p)=[\theta_1(p),\ldots,\theta_M(p)]$, respectively.

Throughout this work, we use model $\theta=[\modelts, \modelnet]$ and hence, operate on object models $\theta(o)=[\modelts(o), \modelnet(o)]$ and prototype models $\theta(p)=[\modelts(p), \modelnet(p)]$, with time-series components $\modelts : \{\varmathbb{R}^T\}$ and network location components $\modelnet : V^+$, where $V$ are the nodes of a network.

\subsection{Similarity functions}
TiCoNE 2 can calculate similarities for object-prototype, object-object, and prototype-prototype pairs using different integrated similarity functions.

\paragraph{Time-series similarity functions.} Firstly, similarities can be calculated based on time-series components $\modelts$. Since each $\modelts$ can contain multiple time-series, time-series similarity functions calculate the average similarity of time-series pairs of the two components. Here, we provide transformations of the Pearson Correlation $S^P$ (\autoref{eqn:pearson}) and of the Euclidean distance $S^E$ (\autoref{eqn:euclidean}).

\begin{dmath}
\simpc(x,y) = \begin{dcases}
\E_{
\substack{t_x \in \modelts(x), \\ t_y \in \modelts(y)}}
\left[\frac{\rho(t_x,t_y)+1}{2}\right] & 
\begin{nullcases}
if $\sigma(x)>0$ \\ \& $\sigma(y)>0$
\end{nullcases} \\
\mbox{undefined} & \mbox{else}
\end{dcases}
\label{eqn:pearson}
\end{dmath}

\begin{dmath}
\simed(x,y) = 
 \E_{
\substack{t_x \in \modelts(x), \\ t_y \in \modelts(y)}}
\left[\frac{m_E -||\modelts(x)-\modelts(y)||}{m_E}\right]
\label{eqn:euclidean}
\end{dmath}
 Similarities of $\simed$ are normalized into the interval $[0,1]$ using the constant $m_E$ which corresponds to the largest observable Euclidean Distance for a given time-series data set $X$. We calculate the constant as $m_E = T \cdot (max(X)-min(X))$, where $max(X)$ and $min(X)$ are the largest and smallest values in any time-series in $X$ and $T$ is the number of their timepoints.

\paragraph{Network location similarity functions.} Secondly, TiCoNE 2 can calculate similarities based on network location components $\modelnet$.
Here, we incorporated a similarity function $\simsp$ (Equation~\ref{eqn:shortestpath}) based on the inversion of the length of the shortest path $|SP(n_1,n_2)|$ in a network $G=(V,E)$ between two node sets $n_1,n_2 \subseteq V$. 

\begin{equation}\label{eqn:shortestpath}
\simsp(x,y)=
\begin{dcases}
1 - \frac{|SP(\modelnet(x), \modelnet(y))|}{m_{SP}} & 
\begin{nullcases}
\mbox{if x and y} \\ \mbox{connected}
\end{nullcases} \\
\mbox{undefined} & \mbox{else}
\end{dcases}
\end{equation}

Here, similarities are normalized into the interval $[0,1]$ using the constant $m_{SP}=|V|-1$, where $|V|$ is the number of nodes in the network. The constant $m_{SP}$ is an upper bound for the shortest path between two nodes in a network with $|V|$ nodes.

\paragraph{Composite similarity functions.} Finally, together with the composite clustering model, we introduce composite similarity functions (\autoref{eqn:composite_sf}) which combine weighted similarities of scaled non-composite similarity functions $\hat{S}_i(x,y)=t_i(S_i(x,y))$.

\begin{equation}\label{eqn:composite_sf}
CS(x,y;\theta)=\frac{\sum_i w_i \cdot S_i'(\theta_i(x),\theta_i(y))}{\sum_i w_i}
\end{equation}

Here weights $w_i$ can be user defined. Furthermore, we scale each $S_i$ using a scaler $t_i$ based on a data-set and similarity-function specific random sample of similarity values. Performing this additional transformation ensures that all child similarity functions are equally contributing to each overall composite similarity value. This is done additionally to the normalization to the interval $[0,1]$ of each respective similarity function, because the functions can produce very different data value distributions in a data-set specific manner. For instance, $\simsp$ highly depends on the overall edge density of the network, and hence normalizing it to the interval $[0,1]$ using the $m_{SP}=|V|-1$ constant, would still result in similarity values with a bias towards low similarities on sparse networks.

A composite similarity function $CS_\theta$ is tied to a clustering model $\theta$, i.e. it requires arguments following its model and its child similarity functions $S_i$ have to be compatible to $\theta_i$.

In this work we make use of two unweighted composite similarity functions $\simpcsp$ and $\simedsp$, each integrating one scaled time-series and one scaled network location similarity function (\Autoref{eqn:composite_psp,eqn:composite_esp}).

\begin{dmath}\label{eqn:composite_psp}
CS^{PC+SP}(x,y) = \frac{1}{2} \cdot \Big[\hat{S}^{PC}(\modelts(x),\modelts(y)) + \hat{S}^{SP}(\modelnet(x),\modelnet(y))\Big]
\end{dmath}

\begin{dmath}\label{eqn:composite_esp}
CS^{ED+SP}(x,y) = \frac{1}{2} \cdot \Big[ \hat{S}^{ED}(\modelts(x),\modelts(y))
+ \hat{S}^{SP}(\modelnet(x),\modelnet(y)) \Big]
\end{dmath}

\subsubsection{Missing Similarities}
Some of TiCoNE's similarity functions are not defined for all inputs. For instance, $\simpc(x,y)$ is only defined for $x,y$ with $\sigma(x) > 0$ and $\sigma(y) > 0$. Likewise, $S^{SP}$ is only defined for pairs of nodes that are connected in the network.

While TiCoNE can handle missing similarity values in different ways, in this work we regarded all composite values themselves as missing if one or more of their child values were missing values. This is more frequently the case, for networks with low edge density and a consequently large number of disconnected node pairs.

If a similarity value is missing, it is ignored in the clustering process and TiCoNE aims at identifying a clustering with the remaining non-missing values.
 
\subsection{Cluster aggregation functions}
Whenever a prototype should be derived for the members of a cluster, several aggregation functions are being used. Specifically, if the clustering model is $\theta$, we use a vector of component specific aggregation functions $A_\theta=[A_{\theta_1},\ldots,A_{\theta_M}]$. Hence, we employ different aggregation functions to derive time-series components $\modelts$ and network location components $\modelnet$ from the cluster members.

While TiCoNE 2 integrates several more choices, in this work we use the following aggregation functions.
\paragraph{Average time series.} Given a cluster $c$. Let $TS(c)$ be the set of all time-series $t_i=(t_{i1},\ldots,t_{iT})$ of members in $c$. We define the average time-series of a cluster as
 $$A^\mu_{\modelts}(c)=\frac{1}{|TS(c)|}\sum_{t_i \in TS(c)}t_i$$
\paragraph{Medoid cluster node.} Given a cluster $c=\{o_i\}$, where members $o_i$ are mapped to nodes in a network $G=(V,E)$. Let $CC(G, c)$ be a partitioning of the members of $c$ into connected components as induced by $G$ and let $cc^*=\argmax_{cc \in CC(G, c)} |cc|$ be the largest such connected component. We define the medoid cluster node of $c$ as 
$$A^{med}_{\modelnet}(G, c) = \argmax_{o_i \in cc^*}\sum_{o_j \neq o_i \in cc^*} S_{net}(o_i, o_j)$$

\subsection{Iterative clustering optimization}

TiCoNE implements a greedy, iterative optimization scheme, in which the following two steps are repeated alternatingly to find (locally) optimal clustering solutions, given a clustering model $\theta$:

\begin{enumerate}
\item Assign objects $o$ to the most similar prototype $p$ as defined by a chosen similarity function $S_\theta$.
\item Refine models $\theta(p)$ for each prototype $p$ according to the members of its cluster using similarity function $S_\theta$ and cluster aggregation functions $A_\theta$.
\end{enumerate}

This leads to clusterings for each of the performed iterations $i \in \{1,\ldots,F\}$, where $C_i$ denotes the clustering of iteration $i$. We denote $C_1$ as the initial clustering, and $C_F$ as the final clustering.

\subsection{Cluster p-values}
TiCoNE 2 allows to calculate empirical, permutation-based p-values for clusters $c$ of a clustering $C$, using one of multiple available permutation functions for time-series data and networks. Here, a fitness function $FS$ is evaluated and compared for each original and permuted cluster. A small p-value for cluster $c$ implies, that few permuted clusters have a fitness score as large as $FS(c)$.

The user can choose the features of a cluster $c$ that should be incorporated into the fitness score from a set of provided features, such as a the object-prototype similarity $F_{OPS}(c;S)=\sum_{o_i \in c} S(o_i,p(c))$, and the number of cluster members $F_N(c)=|c|$.

Furthermore, the user can define a predicate $B: C \times R \mapsto \{\mbox{true}, \mbox{false}\}$ to restrict which random clusters $c_r\in R$ a cluster $c$ should be compared against, where $R$ is the set of all generated random clusters. This is to account for correlations between cluster features and cluster fitness that distort the resulting p-values. For instance, clusters of larger size tend to have smaller average object-prototype similarity.

The p-value for cluster $c$ is then defined as specified in \autoref{eqn:definition-pvalue}.

\begin{equation}\label{eqn:definition-pvalue}
p(c;FS,B) = \frac{\sum_{\{c_r \in R | B(c,c_r)\}} I(FS(c_r)\geq FS(c))}{|\{c_r \in R | B(c,c_r)\}|}
\end{equation}

For this work, we defined the fitness of a cluster $c$ as $FS(c)=F_{OPS}(c;S)$ and chose 

\begin{equation*}
B(c,c_r) = \begin{cases} true & \mbox{if } F_N(c_r) = F_N(c) \\
false & \mbox{else} \end{cases}
\end{equation*}
Hence, we define the fitness of a cluster in terms of its object-prototype similarity, and only compare the fitness of a cluster against the fitness of random clusters of the same size to avoid unfair comparisons. Consequently, we approximate a distinct fitness score distribution for each cluster size.






\subsubsection{Generating functions for random data}\label{sec:permutation-functions}
\paragraph{Random time series data set (R1).}
Given a set of time-series $X =\{t_1,\ldots,t_n\}$, let $min(X)$ and $max(X)$ be the minimal and maximal values in $X$ respectively.
For each input time-series $t_i = (t_{i1},\ldots,t_{iT})$, generate a random time-series $t^*_i = (R_{i1},\ldots,R_{iT})$, with uniformly distributed random variables $R_{ij} \sim U([min(X),max(X)])$.

\paragraph{Shuffle time series individually (R2).}
Given a set of time-series $X =\{t_1,\ldots,t_n\}$.
For each input time-series $t_i = (t_{i1},\ldots,t_{iT})$, generate a random time-series $t^*_i = (R_{i1},\ldots,R_{iT})$, with uniformly distributed random variables $R_{ij} \sim U(\{t_{i1},\ldots,t_{iT}\})$
\paragraph{Shuffle time series data set (R3).}
Given a set of time-series $X =\{t_1,\ldots,t_n\}$, with time-series $t_i=(t_{i1},\ldots,t_{iT})$. Let $V = \cup t_{ij}$ be the set of all values of all time-series of $X$.
For each input time-series $t_i$, generate a random time-series $t^*_i = (R_{i1},\ldots,R_{iT})$, with uniformly distributed random variables $R_{ij} \sim U(V)$.

\paragraph{Random undirected network (R4).}
Given an undirected network $G=(V,E)$ with vertices $V$ and edges $E$. Generate a random network $G^*=(V,E^*)$ with random edges $e^*_i=(R_{i1},R_{i2}) \in E^*$, where $R_{ij} \sim U(V\times V)$ and $|E^*|=|E|$.

This permutation function generates networks that maintain the following properties of the input network:
\begin{itemize}
\item The number of nodes
\item The number of edges
\end{itemize}

Apart from these, it does not ensure or maintain any other properties of the input network. Since edges are randomly attached to nodes, the network topology as well as the distribution of shortest paths between nodes is random as well.

\paragraph{Edge crossovers (R5).}
Given an undirected network $G=(V,E)$ with vertices $V$ and edges $e_i = \{n_{i1},n_{i2}\} \in E$, where $n_{i1},n_{i2} \in V$ are the nodes connected by edge $e_i$. 
Generate a random network $G^*=(V,E^*)$ as follows.

\begin{enumerate}
	\item let $E^*=E$
	\item Take two random edges $e_x=\{n_{x1},n_{x2}\},e_y=\{n_{y1},n_{y2}\} \in E^*$
	\begin{enumerate}
		\item either let $e_x=\{n_{x1},n_{y2}\}$ and $e_y=\{n_{y1},n_{x2}\}$, or
		\item let  $e_x=\{n_{x1},n_{y1}\}$ and $e_y=\{n_{y2},n_{x2}\}$
	\end{enumerate}
	\item Repeat step 2 sufficiently often
\end{enumerate}

This permutation function generates random networks that maintain the following properties of the input network:
\begin{itemize}
\item The number of nodes
\item The number of edges
\item The node degree distribution
\item The joined node degree distribution
\end{itemize}

\subsection{Cluster validity}\label{sec:cluster-validity}
Throughout this work, we use two validity indices to assign quality scores to derived clusterings: The total object-prototype similarity $OPS$ as an internal validity score (\autoref{eqn:validity-sim}), and the Jaccard index $J$ as an external validity score when a gold standard partitioning $GS$ is given (\autoref{eqn:validity-j}). In the latter, $TP(C;GS)$, $FP(C;GS)$ and $FN(C;GS)$ correspond to the number of object pairs that are clustered together in $C$ and $GS$, clustered together in $C$ but not in $GS$, and not clustered together in $C$ nor in $GS$, respectively.

\begin{equation}\label{eqn:validity-sim}
OPS(C;S) = \sum_{c_i \in C} F_{OPS}(c_i;S)
\end{equation}

\begin{equation}\label{eqn:validity-j}
J(C;GS) = \frac{TP(C;GS)}{TP(C;GS)+FP(C;GS)+FN(C;GS)}
\end{equation}

\subsection{Validation data sets}\label{sec:validation-data}

\begin{table*}
	\begin{minipage}[t]{.55\textwidth}
	\caption{Validation data set types generated in this study. For each such type 25 corresponding time-series and network data sets were generated.}
	\begin{tabular}{|l|p{1.5cm}|p{1.4cm}|p{1.8cm}||p{1cm}|}
		\hline
		& \# Objects & \# planted clusters & planted cluster size & \# Data sets \\
		\hline
		\hline
		V1 & 250 & 3 & 10 & 50 \\
		\hline
		V2 & 250 & 5 & 10 & 50 \\
		\hline
		V3 & 250 & 3 & 25 & 50 \\
		\hline
		V4 & 250 & 5 & 25 & 50 \\
		\hline
	\end{tabular}
	\label{tbl:validation_datasets}
		\end{minipage}
		\begin{minipage}[t]{.44\textwidth}
			\caption{Distorted version of the data set types listed in \autoref{tbl:validation_datasets} were generated with varying levels of noise.}
			\begin{tabular}{|l|p{4cm}|}
				\hline
				Type & Noise levels \\
				\hline
				\hline
				Time-Series & $\sigma \in$ \{0.8, 0.4, 0.2, 0.1, 0.05, 0.0\} \\
				\hline
				Network & $d \in$ \{0.0, 0.001, 0.01, 0.05, 0.1, 0.2\} \\
				\hline
			\end{tabular}
			\label{tbl:validation_datasets_noise}
		\end{minipage}
\end{table*}

We generated a set of time-series data sets with $n=250$ objects each, in which $k_e \cdot s$ objects of $k_e$ enriched clusters are embedded in a background of random objects. Here, we varied the number of enriched clusters $k_e$, and their sizes $s$ (see \autoref{tbl:validation_datasets}). For each of these parameter sets, we generated 25 data sets using different random seeds.

Additionally, for each of the resulting time-series data sets, we also generated networks with a matching cluster enrichment, such that objects with similar time-series were also tightly connected in the corresponding network.

Specifically, we generated time-series $t \in [0,1]^T$, with values following a uniform distribution. We assigned identical time-series to all $s$ objects belonging to the same enriched cluster, and random time-series to background objects. We did not enforce random time-series to be dissimilar to enriched time-series.

Network data sets were generated by adding all possible edges for pairs of nodes within each enriched subnetwork (corresponding to objects in the same enriched cluster), and a lower number of edges to, from and between background objects not belonging to any enriched subnetwork.

\subsubsection{Validation data sets with noise}

For each of the data sets described in the previous section, we also generated time-series and network data sets with varying levels of noise.

For time-series data sets, we added a Gaussian error term  with different values for $\sigma$ (see \autoref{tbl:validation_datasets_noise}) to the time-series of objects in the enriched clusters.

For networks, we varied the background edge density to, from, and between background objects with different values for $d$, while keeping the foreground edge density within enriched clusters at $1.0$ (see \autoref{tbl:validation_datasets_noise}).

\subsection{Validation schemes in this work}
\subsubsection{Clustering and iterative optimization}

\begin{table*}
\caption{Parameter values used for performing TiCoNE's implementation of CLARA.}
\centering
\begin{tabular}{|l|l|}
\hline
Parameter & Values \\
\hline
\hline
k & \{10,30\} \\
\hline
CLARA Samples & 1 \\
\hline
CLARA nstart & 50 \\
\hline
CLARA max swaps & 250 \\
\hline
CLARA Swap probability & 0.2 \\
\hline
\end{tabular}
\label{tbl:clustering_params}
\end{table*}

TiCoNE's clustering procedures should be able to recover planted clusters from otherwise random data. We thus clustered the data sets previously described in \autoref{sec:validation-data} using the parameters shown in \autoref{tbl:clustering_params} into $k\in \{10, 30\}$ clusters.

While TiCoNE's implementation of CLARA was used to identify an initial clustering only, we used our iterative optimization procedure to improve it. We executed CLARA with $nstart=50$, each start corresponding to different random initial cluster medoids. Setting this parameter sufficiently large is relevant for coverage of a larger search space, but increases runtime. Hence, we chose 50 as a good compromise.

We applied TiCoNE 2 to (1) the time-series data sets using similarity functions $\simpc$ and $\simed$, the (2) network data sets using $\simsp$, and (3) corresponding time-series and network data sets simultaneously using $\simedsp$ and $\simpcsp$.

For each derived clustering, we evaluated its total object-prototype similarity as an internal cluster validity index, as well as the Jaccard Index as an external index. For the latter, we used the previously known structure of the generated data sets with enriched clusters and background as a gold standard and ignored random objects.

\subsubsection{Cluster P-Values}

We validated TiCoNE's cluster p-values for two different types of data sets and clusterings: (1) clusters of random objects, generated without any enrichment and (2) foreground as well as random background clusters derived from data sets, which contained varying numbers of planted enriched clusters.

\paragraph{Random clusters on random data.}
P-values should be uniformly distributed under $H_0$. We validated this for TiCoNE's cluster p-values using the following procedure.

\begin{enumerate}
\item First, we generate a sufficiently large number (50) of random time-series and network data sets with $n=250$ objects, that do not contain any cluster structure.
\item For each compatible similarity function $S$ and permutation function $R$:
\begin{enumerate}
\item Cluster the data sets using $S$.
\item Calculate cluster p-values using $R$ with 1000 permutations.
\end{enumerate}
\item Thereby, derive one p-value distribution for each combination of similarity function and permutation function.
\item For each of these p-value distributions we evaluate whether they follow a uniform distribution.

\end{enumerate}

\paragraph{Foreground and background clusters on validation data.}
A second validation for TiCoNE's p-values is, that enriched time-series patterns artificially planted in otherwise noisy random data should be assigned a significant p-value.

\begin{enumerate}
	\item Generate validation data sets of types V1-V4 as described previously. 
\item For each compatible similarity function $S$ and permutation function $R$:
\begin{enumerate}
	\item Cluster the generated data sets using $S$. If the total number of clusters in a data set $X$ was $k$ and the number of planted clusters was $k_e$, then we clustered the random objects of $X$ into $k-k_e$ background clusters, and added $k_e$ planted clusters with their respective foreground objects. This way, we ensured that the enriched clusters were part of the clustering, but the remaining objects were clustered with some variation.
	\item Calculate cluster p-values using $R$ with 1000 permutations.
\end{enumerate}
\item Thereby, derive two p-value distributions for each combination of similarity function and permutation function: one distribution for foreground clusters and another one for background clusters.
\item Finally, we evaluate whether foreground cluster p-values are small, background cluster p-values are larger, and whether there was a reasonably large gap between the two distributions.
\end{enumerate}

\section{Results}

Throughout this section, we denote the change in performances in terms of Jaccard index $J$ and object-prototype similarity $OCS$ as $\changej$ and $\changeops$ respectively.

\subsection{TiCoNE recovers planted clusters}

\begin{figure*}
\centering
\includegraphics[width=.9\linewidth]{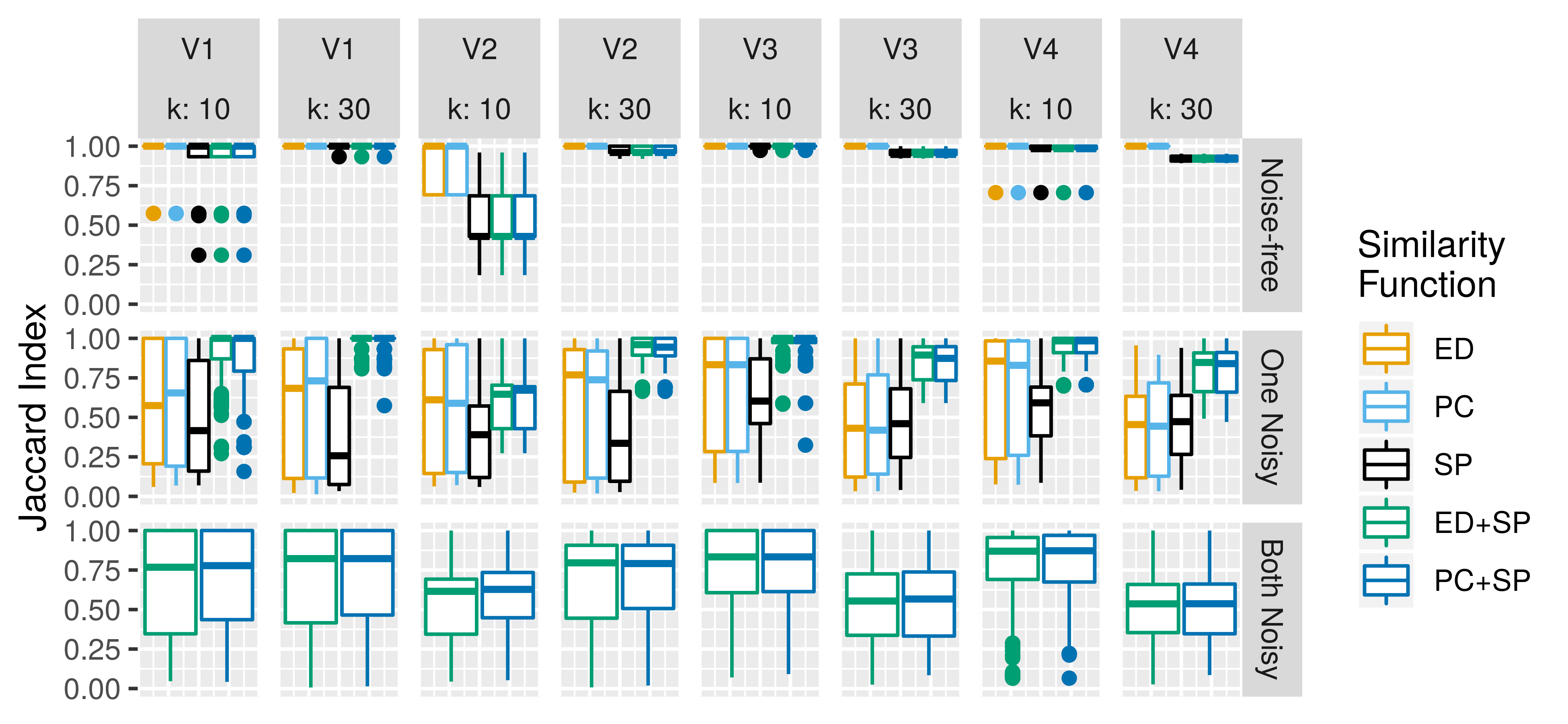}
\caption{Best performances in terms of the Jaccard Index when clustering validation data sets of types V1-V4 using different similarity functions. Non-composite similarity functions $\simed$ (ED), $\simpc$ (PC), and $\simsp$ (SP) are applied to either (1) one noise-free, or (2) one noisy data set at a time, while composite similarity functions $\simedsp$ (ED+SP) and $\simpcsp$ (PC+SP) are applied to either (1) two noise-free, (2) one noise-free and one noisy, or (3) two noisy data sets simultaneously. Boxes summarise achieved performances on 25 generated data sets of the same type.}
\label{fig:bestperformances_simfuncs}
\end{figure*}

\begin{figure*}
\centering
\begin{subfigure}{.58\linewidth}
	\caption{}\label{fig:robustness-timeseries}
\includegraphics[width=1\linewidth]{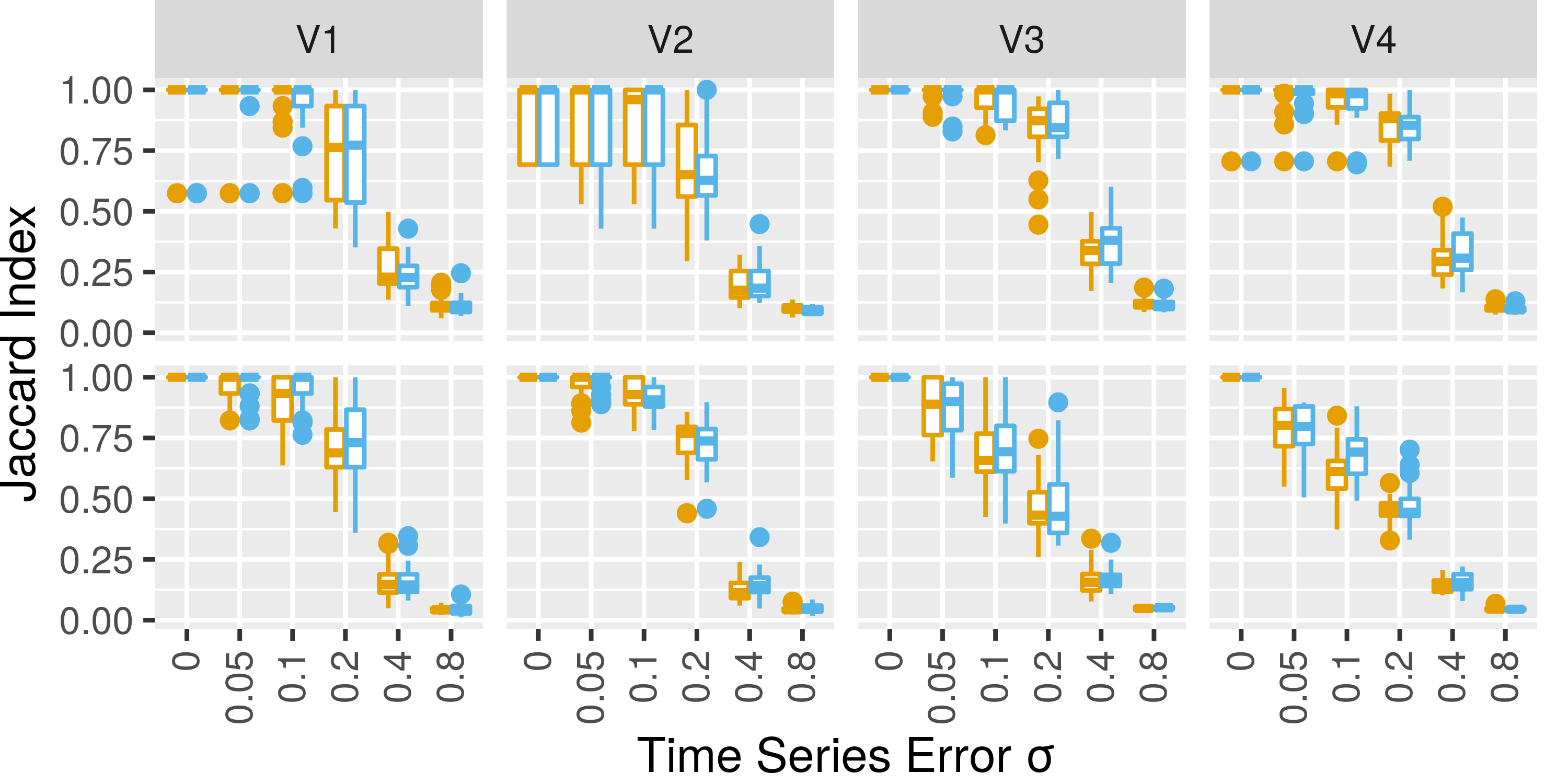}
\end{subfigure}
\begin{subfigure}{.41\linewidth}
	\caption{}\label{fig:robustness-shortestpath}
\includegraphics[width=1\linewidth]{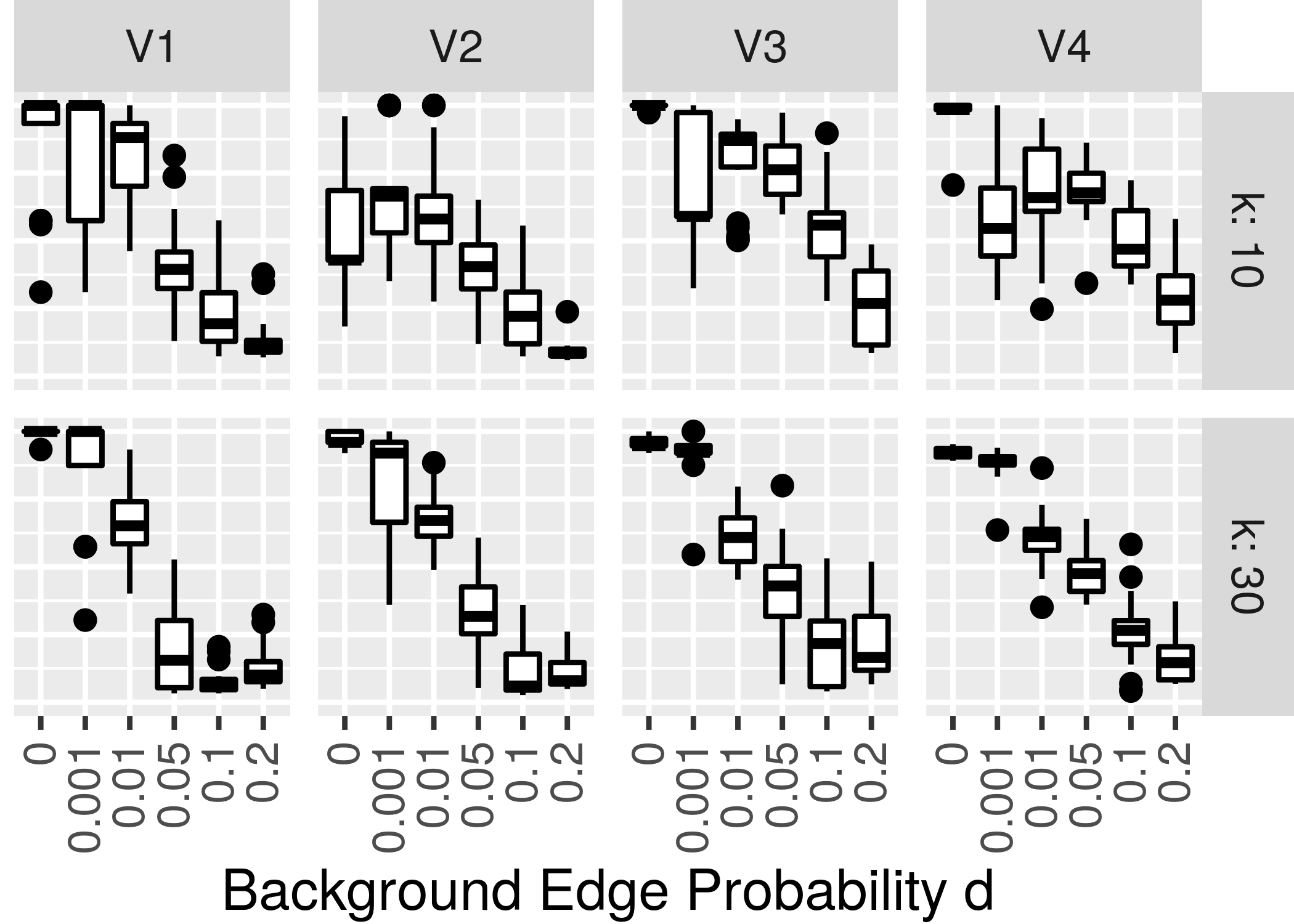}
\end{subfigure}
~
\begin{subfigure}{\linewidth}
\caption{}\label{fig:robustness_composite}
\includegraphics[width=1\linewidth]{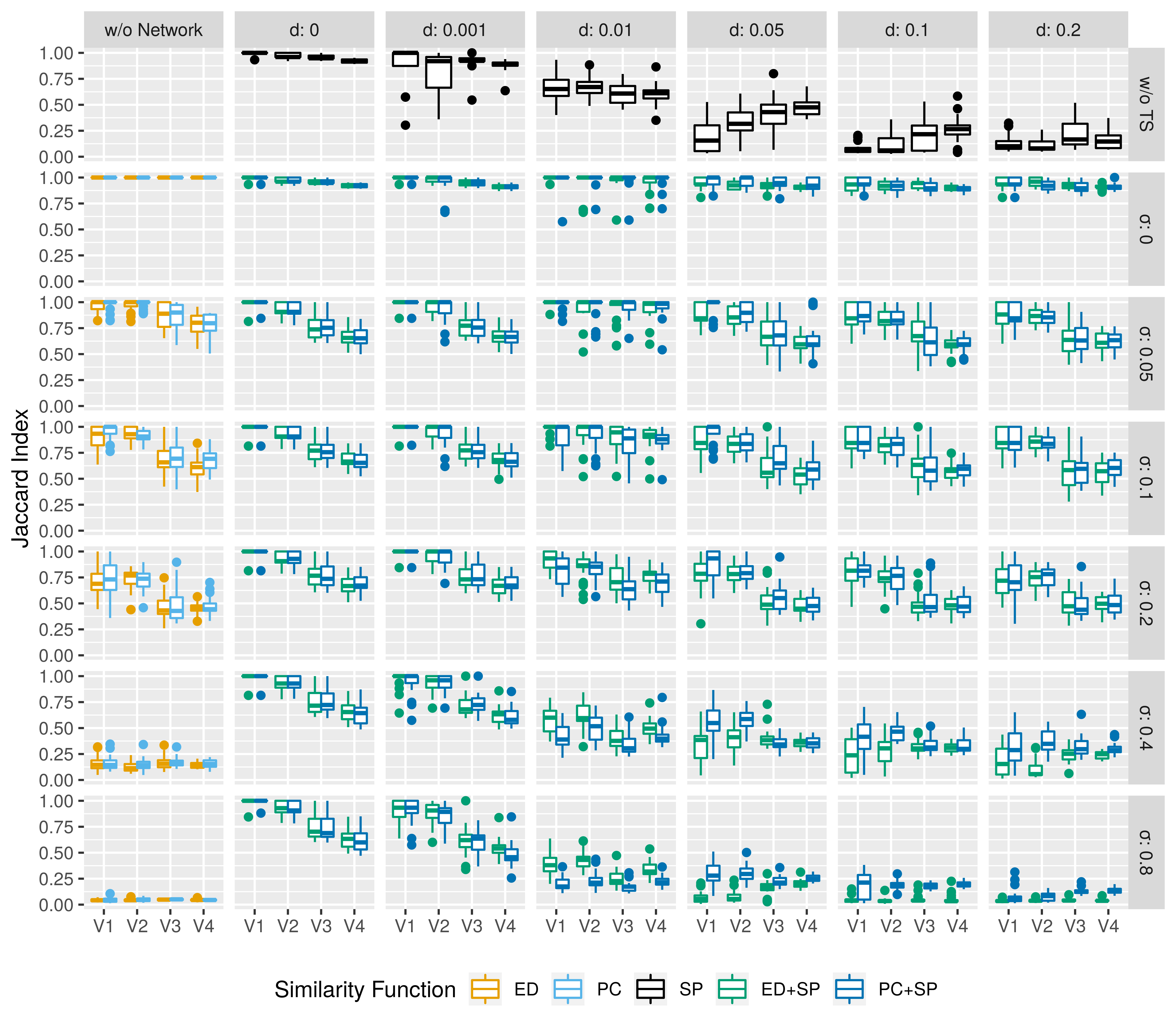}
\end{subfigure}
\caption{Comparison of best achieved performances when clustering data sets of types V1-V4 with varying levels of noise using different similarity functions. Achieved performances when clustering \subref{fig:robustness-timeseries} time-series data with varying levels of noise ($\sigma$) using time-series similarity functions $\simed$ (ED) and $\simpc$ (PC) into $k=10/30$ clusters, \subref{fig:robustness-shortestpath} network data with varying levels of noise ($d$) using shortest path similarity function $\simsp$ (SP) into $k=10/30$ clusters, and \subref{fig:robustness_composite} conjointly clustering time-series and network data with varying levels of noise ($\sigma$ and $d$) with composite similarity functions $\simedsp$ (ED+SP) and $\simpcsp$ (PC+SP).}
\end{figure*}

\subsubsection{Cluster recovery from noise-free data}

All similarity functions perform comparably well on the non-noisy validation data sets (see \autoref{fig:bestperformances_simfuncs}).

In almost all settings, planted clusters can be recovered close to perfectly as indicated by the Jaccard indices varying around 1.0.  Generally, all similarity functions perform comparable, including non-composite as well as composite ones. Clustering data sets of type V2 (5 planted clusters, with 10 members each) into 10 clusters is the most challenging setting. Here, performances of all similarity functions drop, due to a comparably low number of repetitions ($nstart=50$), combined with a low probability to reliably place one medoid out of ten into each of the five relatively small planted clusters with 10 members each.
Here, especially performances of $\simsp$, $\simpcsp$, and $\simedsp$ are affected. This is, because their induced clustering solution search space is highly fragmented for sparse networks, as disconnected nodes cannot be assigned to the same cluster. For instance, if an isolated network node is chosen as a seed medoid, its cluster cannot be assigned any other nodes. Consequently, a poor choice of seed random cluster medoids has a higher impact on the final outcome than for time-series components $\modelts$.

\begin{figure}
\centering
\includegraphics[width=.7\linewidth]{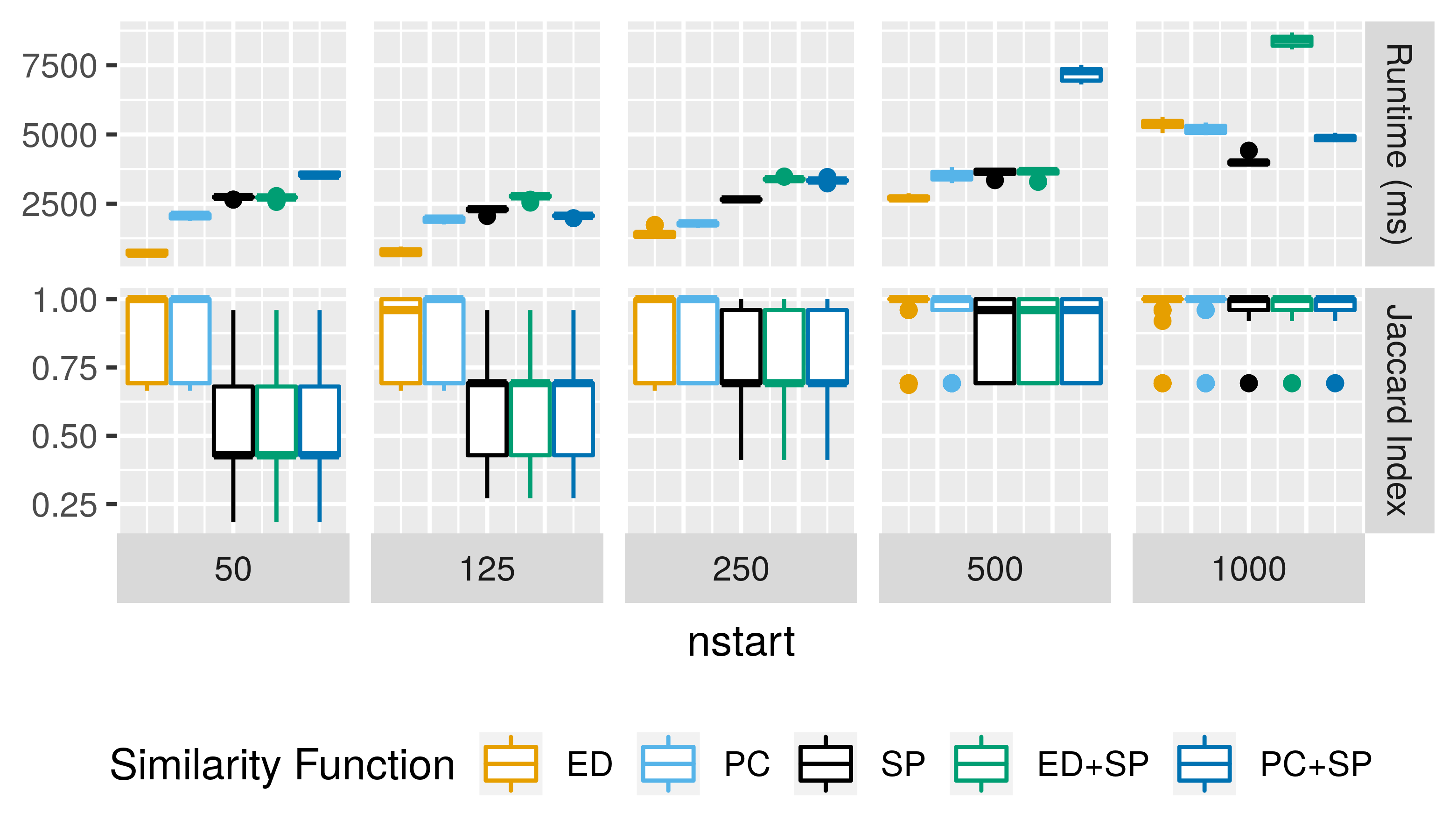}
\caption{Achieved performances and required runtime on the 25 validation data sets of type V2 for increasing values of CLARA's $nstart$ parameter.}
\label{fig:bestperformances_nstarts}
\end{figure}

As can be seen in \autoref{fig:bestperformances_nstarts}, increasing the $nstart$ parameter improves performance on V2 data sets when clustered into $k=10$ clusters, while also increasing average running time. 

\subsubsection{Cluster recovery from noisy data}
As expected, overall clustering of noisy data performs worse in terms of $J$ relative to clustering of noise-free data. However, the drop in cluster performance for non-composite similarity functions is larger than for composite ones for comparable noise levels (see \Autoref{fig:bestperformances_simfuncs}).

Similarity functions $\simed$ and $\simpc$ are comparable in their robustness to noise in the data (\Autoref{fig:bestperformances_simfuncs,fig:robustness_composite}), being more robust when clustering into $k=10$ rather than $k=30$ clusters. When clustering into $k = 10$ clusters, their performances are mostly unaffected for $\sigma \in \{0, 0.05, 0.1\}$ and clearly drop for $\sigma \geq 0.2$. For $k=30$, a drop in performance can already be observed for $\sigma \geq 0.05$. This difference in performance is due to the fact that for $k=30$ the chance of placing multiple medoids into the same planted cluster is higher than for $k=10$, as is the required number of probed seed medoids in order to find a good solution. However, we use a fixed $nstart=50$ regardless of $k$.

Clustering performance on network data using the shortest path similarity function $\simsp$ is more susceptible to noise, than when clustering time-serires data using $\simed$ and $\simpc$ (\Autoref{fig:robustness_composite}). Here, we can already observe a larger drop in performance for low background edge densities $d = 0.001$. This is due to the fact, that already few background edges have the potential to reduce the shortest paths between many background and foreground nodes. Hence, the separation between planted clusters and background becomes more blurred.

Intuitively, the performance of composite functions on two data sets should lie between the better and worse achievable individual performances on either data set. The explanation being, that composite performances should be lower-bound by the worse and upper-bound by the better non-composite performance. For example, if data sets $X$ and $Y$ separately enabled average performances of $J=0.6$ and $J=0.8$, we would expect composite performances to lie roughly in the interval $J\in[0.6,0.8]$. We would not expect composite performances to lie significantly below $J < 0.6$ or above $J > 0.8$.

\paragraph{Two data sets with similarly performing noise levels.}\label{para:cluster-recovery-similar-noise}
We investigated, how composite similarity functions performed on two data sets, for which respective non-composite functions performed similarly well. This is, we investigated performances of $\simedsp$ in settings with noise levels $d$ and $\sigma$, where achieved Jaccard indices were comparable for $\simed$ and $\simsp$ on the respective data sets with those noise levels. Equivalently, we investigated $\simpcsp$ for cases, where $J$ was comparable for $\simpc$ and $\simsp$.

\begin{figure}
\centering
\begin{subfigure}{.45\textwidth}
\caption{}
\includegraphics[trim=15 0 0 0, clip, width=1\linewidth]{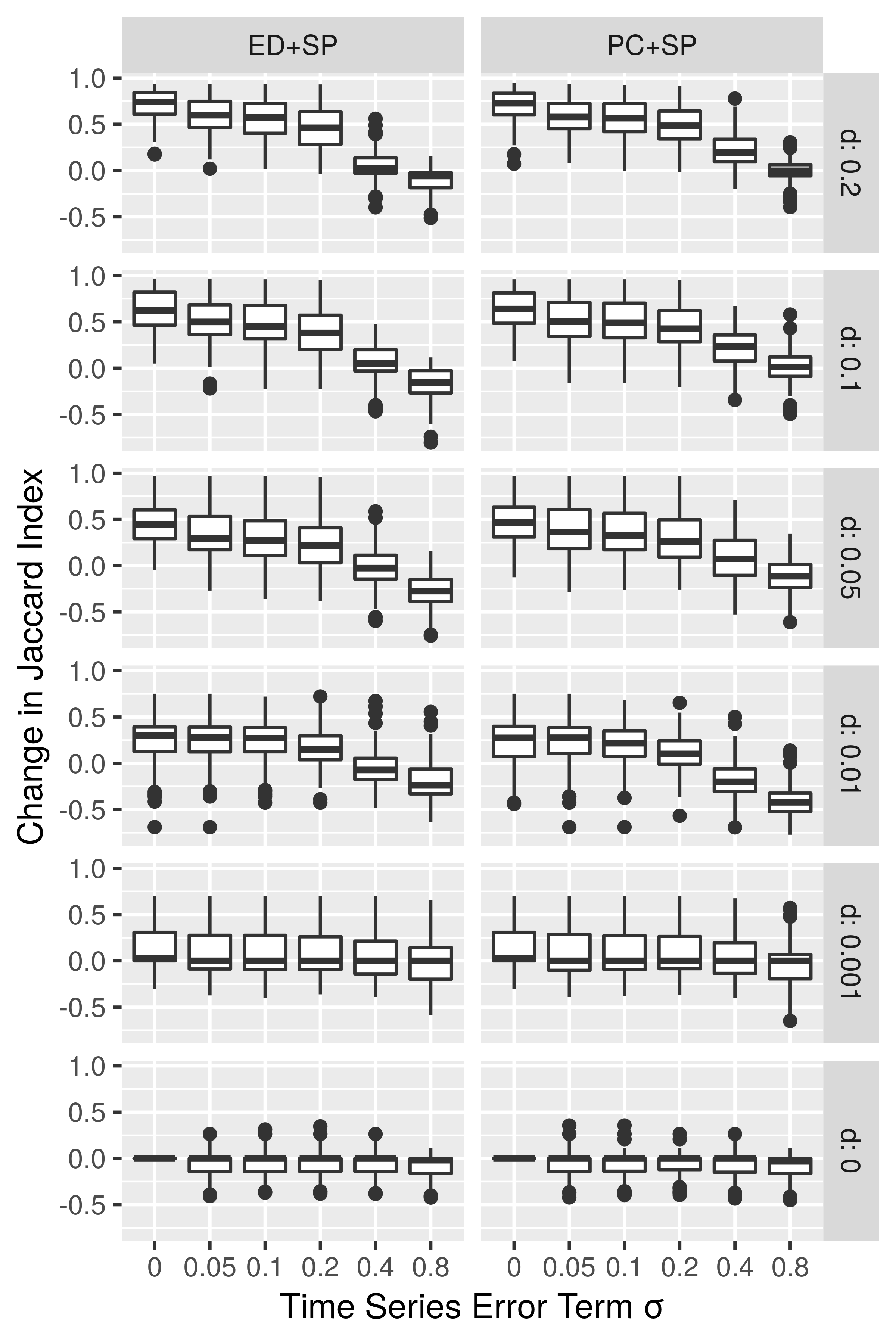}
\label{fig:relative_change_jaccard_composite_vs_timeseries_with_noise}
\end{subfigure}
\begin{subfigure}{.45\textwidth}
\caption{}
\includegraphics[trim=15 0 0 0, clip,width=1\linewidth]{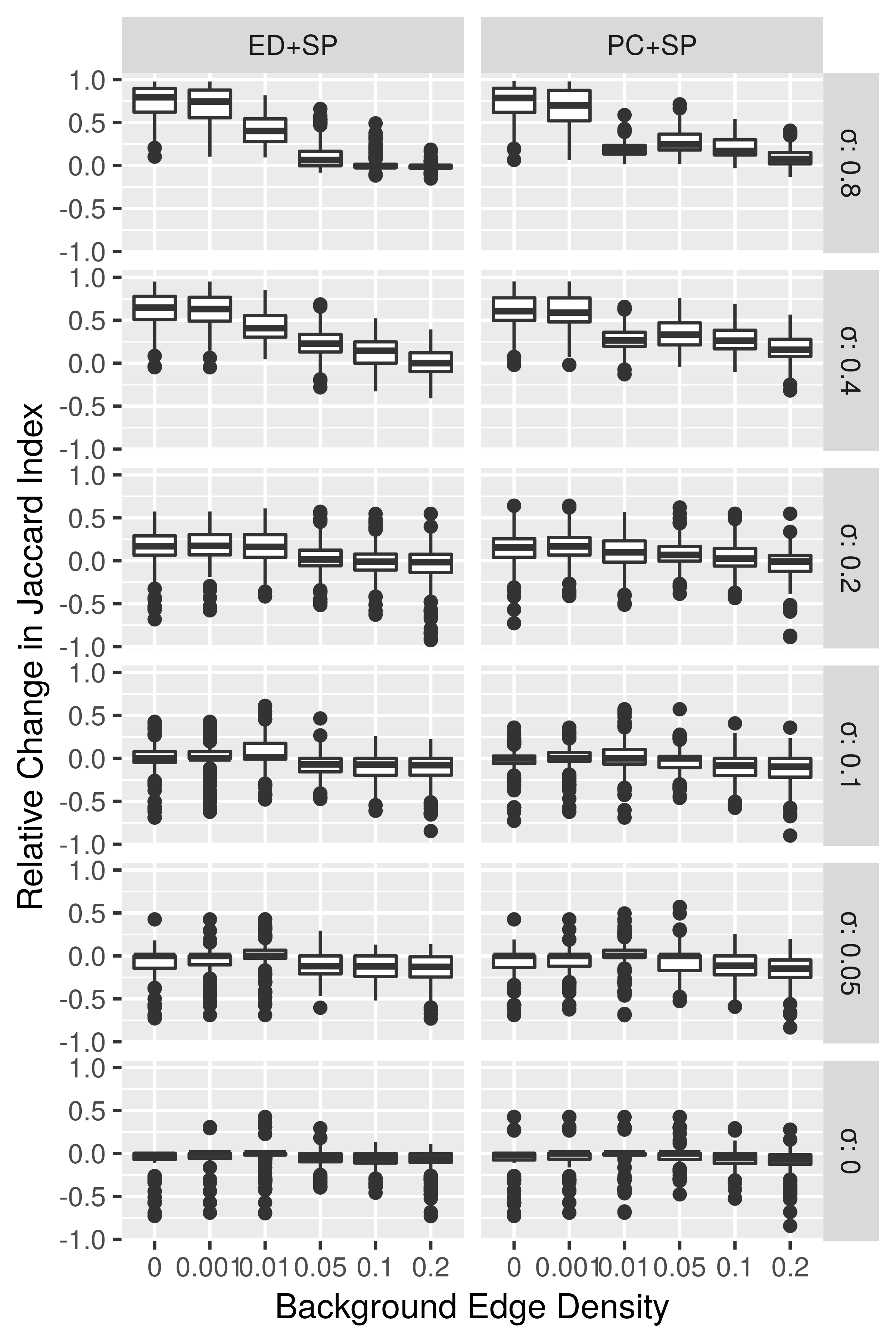}
\label{fig:relative_change_jaccard_composite_vs_shortestpath_with_noise}
\end{subfigure}
\caption{Performances of composite similarity functions $\simedsp$ and $\simpcsp$ in terms of the Jaccard index relative to their corresponding non-composite child similarity functions for varying noise levels in the data. \subref{fig:relative_change_jaccard_composite_vs_timeseries_with_noise} Difference in performance of $\simedsp$ and $\simed$, and of $\simpcsp$ and $\simpc$ for varying noise levels in the time-series data. Values are calculated as $J(\simedsp)-J(\simed)$ and $J(\simpcsp)-J(\simpc)$ respectively. \subref{fig:relative_change_jaccard_composite_vs_shortestpath_with_noise} Relative performances of $\simedsp$ and $\simpcsp$ to $\simsp$ for varying noise levels in the network data. Values are calculated as $J(\simedsp)-J(\simsp)$ and $J(\simpcsp)-J(\simsp)$ respectively.}
\end{figure}

First, we investigated settings where individual performances on either data set were high, i.e. where both noise levels were low. This is the case for $\sigma=0.05$ and $d=0.001$, where $\simed$ and $\simpc$ achieved average performances of $J\in[0.793,0.975]$ and $\simsp$ of $J\in[0.793,0.92]$ across data types V1-V4. Here, both composite similarity functions $\simedsp$ and $\simpcsp$ perform on average slightly better than $\simsp$ on the network data alone with $\changej=+0.078$ and $\changej=+0.075$ respectively (see \autoref{fig:relative_change_jaccard_composite_vs_shortestpath_with_noise}), but slightly worse than $\simed$ and $\simpc$ on the time-series data with $\changej=-0.059$ and $\changej=-0.06$ respectively (see \autoref{fig:relative_change_jaccard_composite_vs_timeseries_with_noise}).

Next, we inspected composite performances when both data sets led to only medium performances. For instance, this is the case for $d=0.01$ and $\sigma=0.1$, where $\simed$ and $\simpc$ achieve average performances of $J\in[0.644,0.93]$ and $\simsp$ of $J\in[0.603,0.674]$. Here, $\simedsp$ and $\simpcsp$ outperform $\simsp$ clearly with $\changej=+0.235$ and $\changej=+0.197$ respectively. Further, they also outperform $\simed$ and $\simpc$ with $\changej=+0.064$ and $\changej=+0.013$.

When both data sets led to poor performances and had high levels of noise, for instance for $d=0.1$ and $\sigma=0.4$, $\simed$ and $\simpc$ achieved average performances of $J\in[0.139,0.168]$ and $\simsp$ of $J\in[0.074,0.255]$. For these data sets, $\simedsp$ performs superior to $\simed$ with $\changej=+0.132$ and $\simpcsp$ performs superior to $\simpc$ with $\changej=+0.275$. Both composite similarity functions perform superior to $\simsp$ with $\changej=+0.07$ and $\changej=+0.225$ respectively.

Only for data sets of highest levels of noise with $d=0.2$ and $\sigma=0.8$, performances of composite functions were inferior to their non-composite counterparts.


\paragraph{High quality network, low quality time-series}
We then focused on cases of two data sets, that led to performances on different levels when analyzed separately.

In particular, we first investigated performances for high quality network data (low $d$), and low quality time-series data (high $\sigma$). Most notably, for $d=0$ varying levels of $\sigma$ do not have much influence on the achieved performances relative to clustering the network data alone. This is, if the network data is noise-free, drop in performance is minimal when conjointly analyzing with noisy time-series data.

When clustering highly noisy time-series data ($\sigma=0.4$) and network data with little noise ($d=0.01$) using $\simedsp$, we observe an average performance increase of $\changej = +0.418$ relative to $\simed$ on the noisy time-series data alone (see \autoref{fig:robustness_composite}). However, at the same time $\simedsp$ performs inferior to $\simsp$ ($\changej = -0.054$) when clustering the high-quality network data alone. This trend is even more prominent, for $\sigma=0.8$ ($\changej=+0.418$ and $\changej=-0.188$). This drop in performance however is inverted if the noise level of the time-series data is not as high, but lower with $\sigma = 0.2$: Then, $\simedsp$ performs superior over both non-composite functions $\simed$ ($\changej=+0.159$) as well as $\simsp$ ($\changej=+0.158$).

Equivalently to $\simedsp$, for the same noise levels $\simpcsp$ leads to an average performance gain of $\changej = +0.277$ relative to $\simpc$, but a drop in performance of $\changej = -0.184$ relative to $\simsp$. Also, if the noise level of the time-series data is $\sigma = 0.2$ instead, then $\simpcsp$ performs superior over both $\simpc$ ($\changej=+0.103$) and $\simsp$ ($\changej=+0.112$).

Generally, composite performances are superior as long as the time-series noise level is not exceeding $\sigma=0.2$. If the noise level of the time-series data is too prominent, it negatively affects overall composite performance even if network quality is high.

\paragraph{High quality time-series, low quality network}

Next we evaluated performances for high quality time series data (low $\sigma$), and low quality network data (high $d$). Also here it can be seen that for noise-free time-series data ($\sigma = 0$) varying levels of $d$ do not have much influence on the achieved performances relative to clustering the time-series data alone. Hence, if the time-series data is noise-free, the drop in performance is minimal when conjointly analyzing with noisy network data.

Furthermore, the same general trend can be observed as for the previously discussed opposite case (high quality network, low quality time-series). For instance, when evaluating composite performances for $d=0.1$ and $\sigma=0.1$, $\simedsp$ achieves an average performance gain of $\changej = +0.482$ relative to $\simsp$ but is inferior to $\simed$ with $\changej = -0.1$. This trend is even more prominent for $d=0.2$ ($\changej=+0.568$ and $\changej=-0.112$). This drop in performance however is inverted if the noise level of the network data is not as high, but lower with $d = 0.01$: Then, $\simedsp$ performs superior over both non-composite functions $\simed$ ($\changej=+0.064$) as well as $\simsp$ ($\changej=+0.235$).

Hence, the general trend here is also that composite functions perform superior as long as the network noise level is not exceeding $d=0.01$. For higher network noise levels, analyzing both data types conjointly may perform inferior to analyzing the high quality data alone.

\subsection{TiCoNE's iterative optimization enhances cluster validity}

\begin{figure*}
\centering
\begin{subfigure}{.75\linewidth}
\caption{}
\includegraphics[trim=0 40 0 0, clip, width=\linewidth]{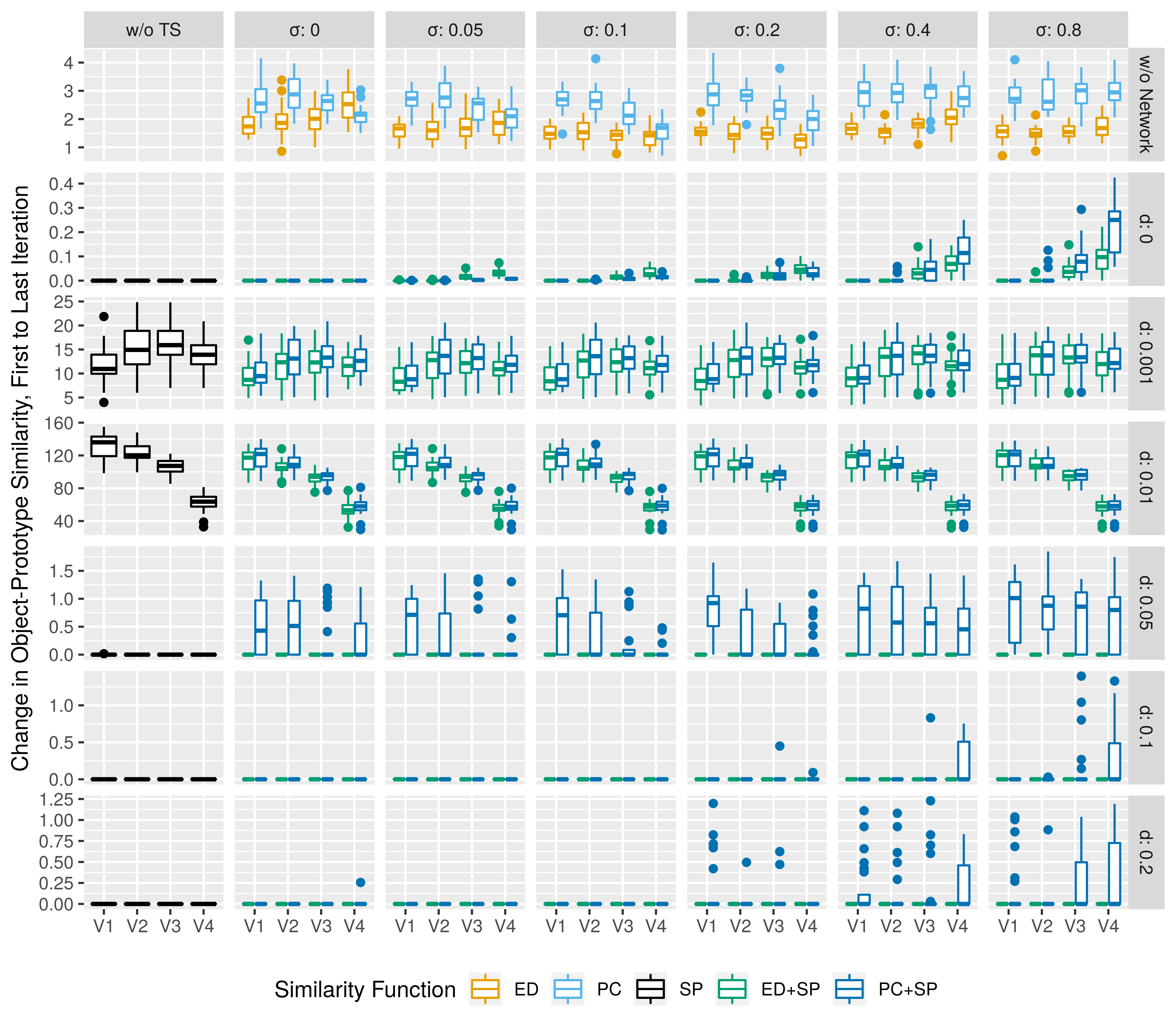}
\label{fig:relative_change_sim_first_to_last}
\end{subfigure}
~
\begin{subfigure}{.75\linewidth}
\caption{}
\includegraphics[width=\linewidth]{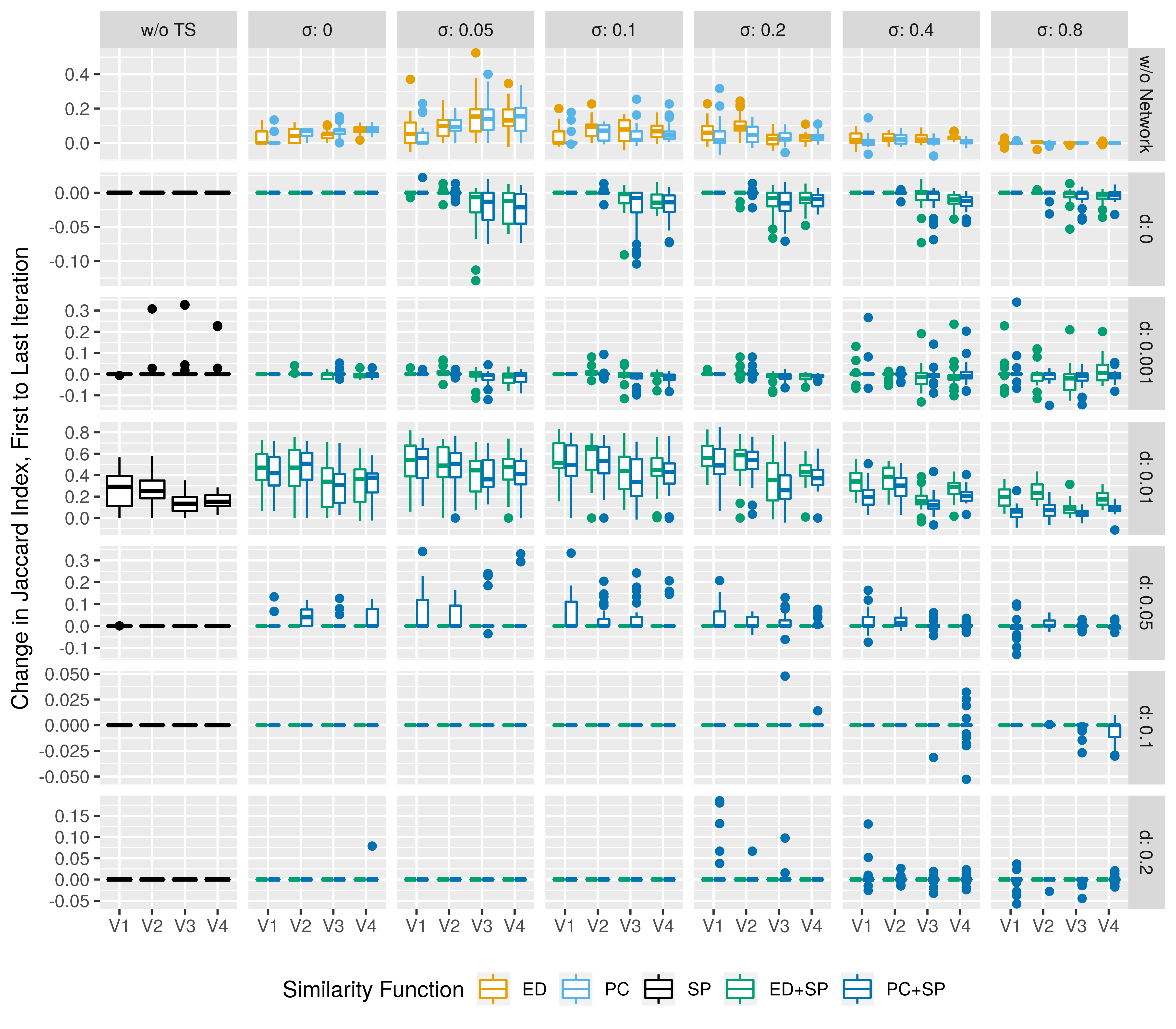}
\label{fig:relative_change_jaccard_first_to_last}
\end{subfigure}
\caption{The changes in \subref{fig:relative_change_sim_first_to_last} object-prototype similarity (internal cluster validity) and in \subref{fig:relative_change_jaccard_first_to_last} Jaccard index (external cluster validity) when applying TiCoNE's iterative optimization procedure using different similarity functions on data set types V1-V4 with differing levels of noise.}
\end{figure*}

For all validation data set types V1-V4, we evaluated the gain in performance in terms of the object-prototype similarity (internal validation) as well as the Jaccard index (external validation) from first to last iteration of TiCoNE's iterative clustering optimization scheme.

Generally, for all evaluated similarity functions object-prototype similarities increase or stay constant from initial to final clustering (see \autoref{fig:relative_change_sim_first_to_last}). For $\simed$ performance gains tend to be higher for lower noise levels, while for $\simpc$ for higher noise levels. Generally, $\simpc$ performances improve more from first to last iteration, indicating that initial clusterings derived using the latter may be less optimal than the ones derived using $\simed$. For the shortest path based similarity function as well as composite functions, the gain is highest for $d=0.01$, medium for $d=0.001$ and close to zero for any other value of $d$. This is due to the network topology induced for different values of $d$ as well as our parameter value choice of $nstart=50$.

When evaluating the performance changes in terms of the Jaccard index, there is an improvement in performance from first to last iteration for most cases, but some data sets also lead to a limited decrease of up to $\changej\leq -0.1$. Such data sets are mostly ones with a high amount of noise, such as for $\sigma\in\{0.4,0.8\}$. Also, interestingly there are more such outliers with decreased performance for $d\in\{0.001,0.01\}$ than for $d>0.01$, which again may be explained by the network topology induced by these particular values for $d$. Furthermore, $\simpcsp$ shows a performance improvement in most cases with $d\leq0.05$. This stands in contrast to $\simedsp$, where performances stay constant for most such data sets.

Overall, TiCoNE's iterative clustering optimization improves not only the object-prototype similarity ($OPS$), but also implicitly the Jaccard index. Note, that this will always be the case, if the gold standard is an accurate description of the internal structure of the data set and its induced object similarities.

\subsection{TiCoNE produces meaningful cluster p-values}
\paragraph{FDR on random data}

\begin{figure*}
\centering
\includegraphics[width=.8\linewidth]{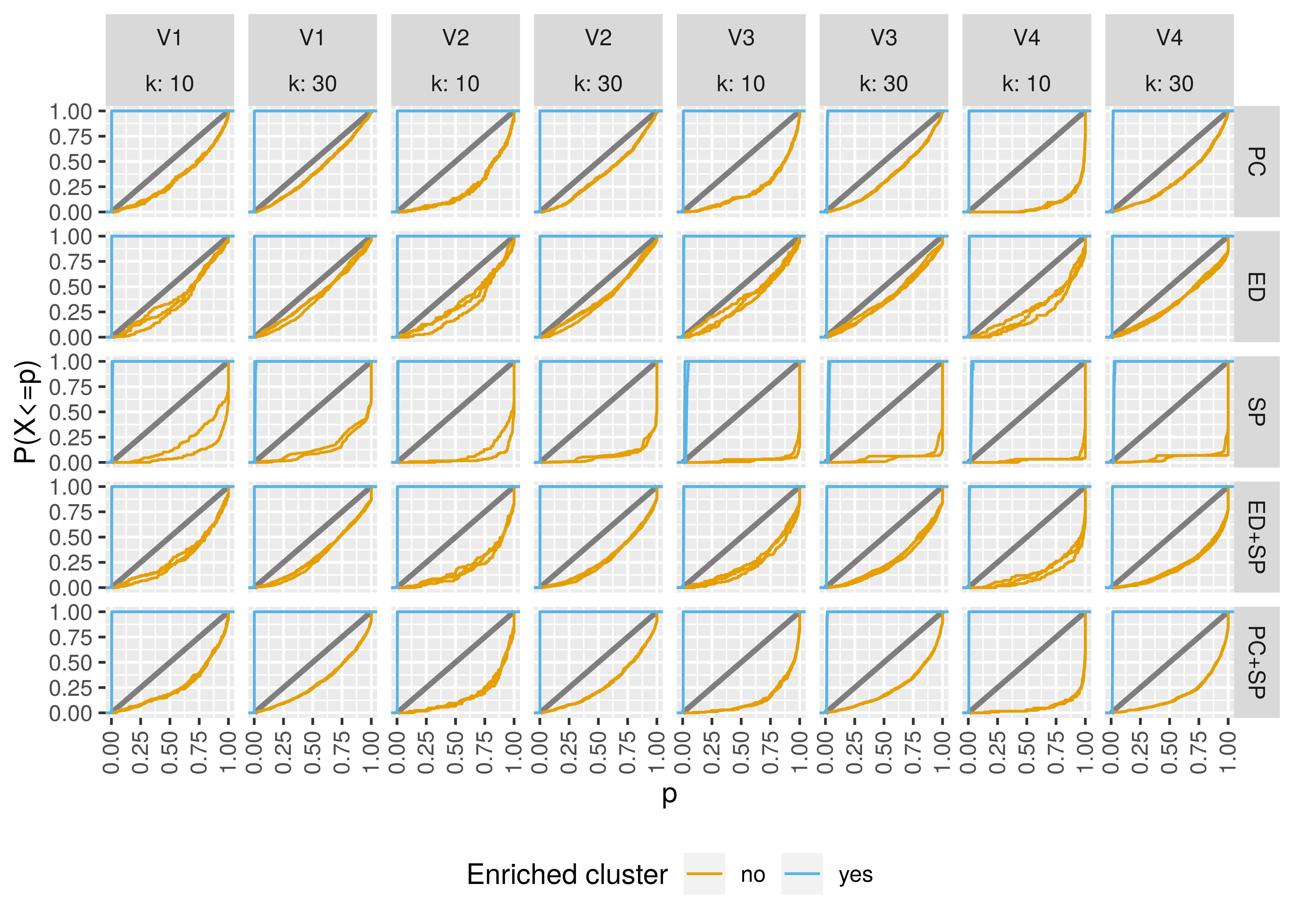}
\caption{Distributions of p-values for enriched and random background clusters derived when clustering different data set types V1-V4 into $k=10/30$ clusters using different similarity functions. Each box contains one foreground and one background cluster p-value distribution for each permutation function. The diagonal corresponds to an ideal uniform distribution over the interval $[0,1]$.}
\label{fig:pvalues-hidden}
\end{figure*}

\begin{table*}
{\footnotesize
\setlength\tabcolsep{3pt} 
\begin{tabular}{|r|p{.85cm}|p{.5cm}|rr|rr|rr|rr|rr|rr|rr|rr|}
  \hline
  \multirow{3}{.25cm}{$k$} & \multirow{3}{.85cm}{$S$} & \multirow{3}{.25cm}{$R$} & \multicolumn{4}{c}{V1} & \multicolumn{4}{c}{V2} & \multicolumn{4}{c}{V3} & \multicolumn{4}{c|}{V4} \\
            \cline{4-19} 
    &  &  & \multicolumn{2}{c|}{$\alpha=0.01$} & \multicolumn{2}{c|}{$\alpha=0.05$} & \multicolumn{2}{c|}{$\alpha=0.01$} & \multicolumn{2}{c|}{$\alpha=0.05$} & \multicolumn{2}{c|}{$\alpha=0.01$} & \multicolumn{2}{c|}{$\alpha=0.05$} & \multicolumn{2}{c|}{$\alpha=0.01$} & \multicolumn{2}{c|}{$\alpha=0.05$} \\ 
    &  &  & FNR & FPR & FNR & FPR & FNR & FPR & FNR & FPR & FNR & FPR & FNR & FPR & FNR & FPR & FNR & FPR \\ 
  \hline
    \multirow{14}{.25cm}{10} & \multirow{3}{.85cm}{ED} & R1 & 0.00 & 0.00 & 0.00 & 0.01 & 0.00 & 0.00 & 0.00 & 0.01 & 0.00 & 0.00 & 0.00 & 0.00 & 0.00 & 0.00 & 0.00 & 0.00 \\ 
     &  & R2 & 0.00 & 0.00 & 0.00 & 0.00 & 0.00 & 0.00 & 0.00 & 0.00 & 0.06 & 0.00 & 0.00 & 0.01 & 0.00 & 0.00 & 0.00 & 0.00 \\ 
     &  & R3 & 0.00 & 0.00 & 0.00 & 0.01 & 0.00 & 0.00 & 0.00 & 0.02 & 0.00 & 0.00 & 0.00 & 0.02 & 0.00 & 0.00 & 0.00 & 0.00 \\ 
          \cline{2-19}
     & \multirow{3}{.85cm}{ED+SP} & R1 & 0.00 & 0.00 & 0.00 & 0.01 & 0.00 & 0.00 & 0.00 & 0.00 & 0.00 & 0.00 & 0.00 & 0.00 & 0.00 & 0.00 & 0.00 & 0.00 \\ 
     &  & R2 & 0.00 & 0.00 & 0.00 & 0.00 & 0.00 & 0.00 & 0.00 & 0.00 & 0.12 & 0.00 & 0.00 & 0.00 & 0.00 & 0.00 & 0.00 & 0.00 \\ 
     &  & R3 & 0.00 & 0.00 & 0.00 & 0.02 & 0.00 & 0.00 & 0.00 & 0.00 & 0.00 & 0.00 & 0.00 & 0.02 & 0.00 & 0.00 & 0.00 & 0.00 \\ 
          \cline{2-19}
     & \multirow{3}{.85cm}{PC} & R1 & 0.00 & 0.00 & 0.00 & 0.01 & 0.00 & 0.00 & 0.00 & 0.00 & 0.00 & 0.00 & 0.00 & 0.00 & 0.00 & 0.00 & 0.00 & 0.00 \\ 
     &  & R2 & 0.00 & 0.00 & 0.00 & 0.01 & 0.00 & 0.00 & 0.00 & 0.00 & 0.00 & 0.00 & 0.00 & 0.00 & 0.00 & 0.00 & 0.00 & 0.00 \\ 
     &  & R3 & 0.00 & 0.00 & 0.00 & 0.01 & 0.00 & 0.00 & 0.00 & 0.00 & 0.00 & 0.00 & 0.00 & 0.00 & 0.00 & 0.00 & 0.00 & 0.00 \\ 
          \cline{2-19}
     & \multirow{3}{.85cm}{PC+SP} & R1 & 0.00 & 0.00 & 0.00 & 0.01 & 0.00 & 0.00 & 0.00 & 0.00 & 0.00 & 0.00 & 0.00 & 0.00 & 0.00 & 0.00 & 0.00 & 0.00 \\ 
     &  & R2 & 0.00 & 0.00 & 0.00 & 0.01 & 0.00 & 0.00 & 0.00 & 0.00 & 0.00 & 0.00 & 0.00 & 0.00 & 0.00 & 0.00 & 0.00 & 0.00 \\ 
     &  & R3 & 0.00 & 0.00 & 0.00 & 0.01 & 0.00 & 0.00 & 0.00 & 0.00 & 0.00 & 0.00 & 0.00 & 0.00 & 0.00 & 0.00 & 0.00 & 0.00 \\ 
          \cline{2-19}
     & \multirow{2}{.85cm}{SP} & R4 & 0.00 & 0.00 & 0.00 & 0.00 & 0.00 & 0.00 & 0.00 & 0.00 & 1.00 & 0.00 & 0.00 & 0.00 & 1.00 & 0.00 & 0.00 & 0.00 \\ 
     &  & R5 & 0.16 & 0.00 & 0.00 & 0.00 & 0.00 & 0.00 & 0.00 & 0.00 & 1.00 & 0.00 & 0.00 & 0.00 & 1.00 & 0.00 & 0.00 & 0.00 \\ 
     \hline
    \multirow{14}{.25cm}{30} & \multirow{3}{.85cm}{ED} & R1 & 0.00 & 0.00 & 0.00 & 0.03 & 0.00 & 0.00 & 0.00 & 0.01 & 0.66 & 0.00 & 0.00 & 0.02 & 0.68 & 0.00 & 0.00 & 0.01 \\ 
     &  & R2 & 0.00 & 0.00 & 0.00 & 0.01 & 0.00 & 0.00 & 0.00 & 0.00 & 0.54 & 0.00 & 0.00 & 0.01 & 0.44 & 0.00 & 0.00 & 0.01 \\ 
     &  & R3 & 0.00 & 0.00 & 0.00 & 0.03 & 0.00 & 0.00 & 0.00 & 0.01 & 0.64 & 0.00 & 0.00 & 0.03 & 0.60 & 0.00 & 0.00 & 0.01 \\ 
     \cline{2-19}
     & \multirow{3}{.85cm}{ED+SP} & R1 & 0.00 & 0.00 & 0.00 & 0.01 & 0.00 & 0.00 & 0.00 & 0.01 & 0.20 & 0.00 & 0.00 & 0.01 & 0.44 & 0.00 & 0.00 & 0.00 \\ 
    &  & R2 & 0.00 & 0.00 & 0.00 & 0.01 & 0.00 & 0.00 & 0.00 & 0.01 & 0.20 & 0.00 & 0.00 & 0.00 & -- & -- & -- & -- \\ 
     &  & R3 & 0.00 & 0.00 & 0.00 & 0.01 & 0.00 & 0.00 & 0.00 & 0.01 & 0.28 & 0.00 & 0.00 & 0.01 & 0.32 & 0.00 & 0.00 & 0.01 \\ 
          \cline{2-19}
     & \multirow{3}{.85cm}{PC} & R1 & 0.00 & 0.00 & 0.00 & 0.02 & 0.00 & 0.00 & 0.00 & 0.01 & 0.98 & 0.00 & 0.00 & 0.01 & 0.98 & 0.00 & 0.00 & 0.01 \\ 
     &  & R2 & 0.00 & 0.00 & 0.00 & 0.01 & 0.00 & 0.00 & 0.00 & 0.01 & 1.00 & 0.00 & 0.00 & 0.01 & 0.94 & 0.00 & 0.00 & 0.01 \\ 
     &  & R3 & 0.00 & 0.00 & 0.00 & 0.02 & 0.00 & 0.00 & 0.00 & 0.01 & 1.00 & 0.00 & 0.00 & 0.01 & 0.96 & 0.00 & 0.00 & 0.01 \\ 
          \cline{2-19}
     & \multirow{3}{.85cm}{PC+SP} & R1 & 0.00 & 0.00 & 0.00 & 0.02 & 0.00 & 0.00 & 0.00 & 0.01 & 0.72 & 0.00 & 0.00 & 0.01 & 0.72 & 0.00 & 0.00 & 0.01 \\ 
     &  & R2 & 0.00 & 0.00 & 0.00 & 0.02 & 0.00 & 0.00 & 0.00 & 0.01 & 0.84 & 0.00 & 0.00 & 0.01 & 0.84 & 0.00 & 0.00 & 0.01 \\ 
     &  & R3 & 0.00 & 0.00 & 0.00 & 0.02 & 0.00 & 0.00 & 0.00 & 0.01 & 0.88 & 0.00 & 0.00 & 0.01 & 0.88 & 0.00 & 0.00 & 0.01 \\ 
          \cline{2-19}
     & \multirow{2}{.85cm}{SP} & R4 & 0.00 & 0.00 & 0.00 & 0.00 & 0.00 & 0.00 & 0.00 & 0.00 & 1.00 & 0.00 & 0.00 & 0.00 & 1.00 & 0.00 & 0.00 & 0.00 \\ 
     &  & R5 & 0.20 & 0.00 & 0.00 & 0.00 & 0.00 & 0.00 & 0.00 & 0.00 & 1.00 & 0.00 & 0.00 & 0.00 & 1.00 & 0.00 & 0.00 & 0.00 \\ 
   \hline
\end{tabular}
}
\caption{False positive rates (FPR) and false negative rates (FNR) of cluster p-values calculated using different permutation functions R1-R5. Clusters are derived from validation data sets of the different types V1-V4 using different similarity functions. Error rates are relative to the significance level $\alpha=0.01$ or $\alpha=0.05$.}
\label{tbl:error-rates-hidden}
\end{table*}

\begin{figure*}
\centering
\includegraphics[width=.6\linewidth]{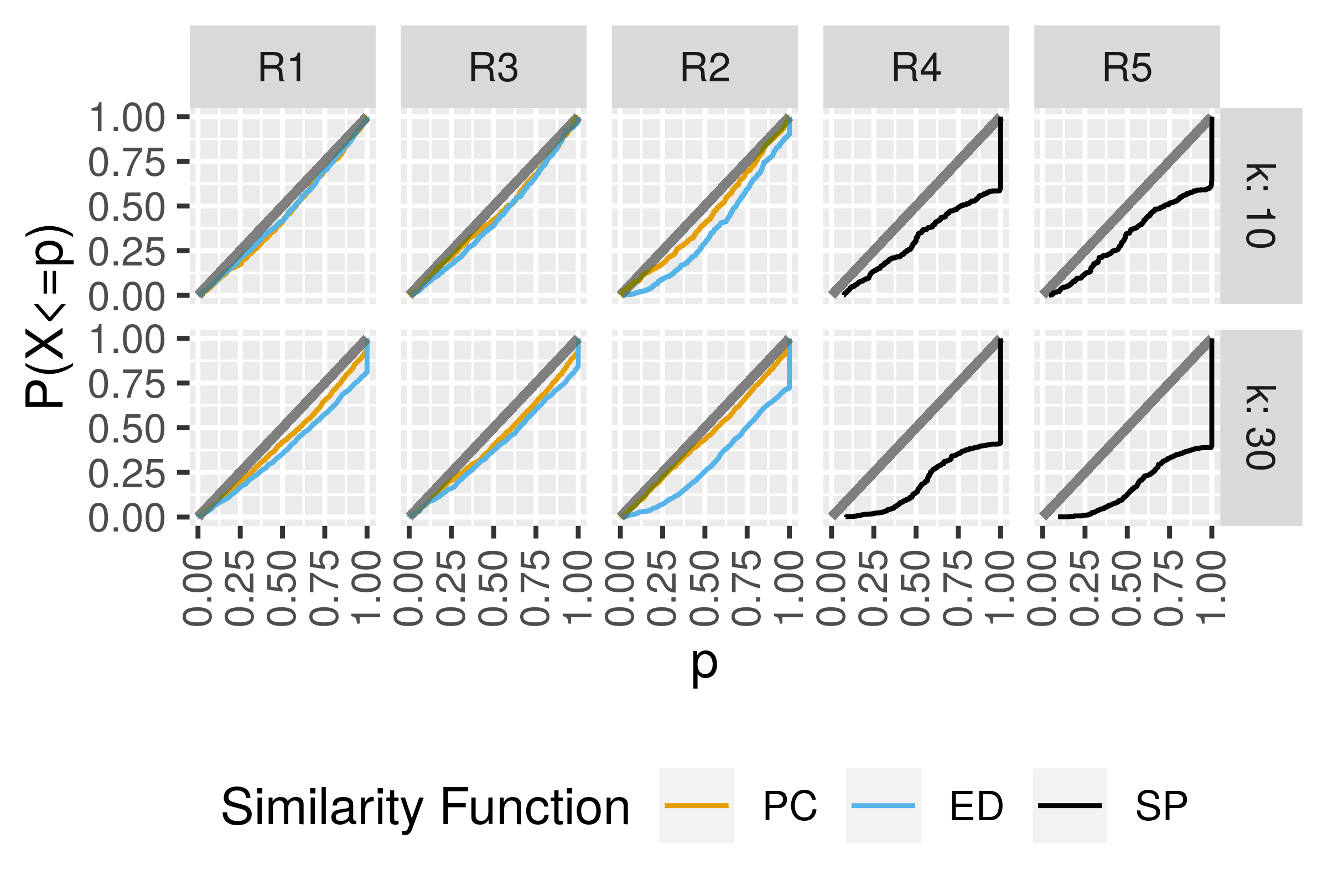}
\caption{Distributions of cluster p-values calculated using different permutation functions. Clusters are derived from randomly generated data sets when clustered into $k=10$ and $k=30$ using different similarity functions. The diagonal corresponds to an ideal uniform distribution over the interval $[0,1]$.}
\label{fig:pvalues-uniform}
\end{figure*}

As can be seen in \autoref{fig:pvalues-uniform}, when clustering random data without enrichment into $k=10/30$ clusters and calculating p-values for clusters with the time-series data permutation functions R1,R2,R3, all resulting p-value distributions were very close to the expected uniform distribution. In the ideal case, the empirical cumulative distribution function (ECDF) of each p-value distribution resembles a perfect line from (0,0) to (1,1). This corresponds to the fact, that p-values should be uniformly distributed under the null hypothesis $H_0$, i.e. the cluster's object-cluster similarity is not significantly better than expected for a random cluster of equal size. Notably, all three permutation functions succeed in sufficiently controlling the type 1 error rate for each fixed $\alpha$, i.e. $\forall \alpha: P(H_1|H_0) \leq \alpha$.

Similarly, both network data permutation functions $R4$ and $R5$ facilitate close to uniformly distributed p-value distributions for $k=10$ and $k=30$. However, for both of them there is a bias towards large p-values close to $p=1.0$. This bias stems from a large number of clusters with very few members in the generated random clusterings. Since we are dealing with unweighted networks, there are limited possibilities for constellations of object-cluster similarities in small clusters. For instance, for a cluster $c_i$ with two members $c_i=\{o_x,o_y\}$ it always holds that $|SP(o_x,o_y)|=1$, and hence $\simsp(o_x,o_y)=1-\frac{1}{m_{SP}}$. Since one of the two has to be the cluster prototype, it also follows that $F_{OPS}=1-\frac{1}{m_{SP}}$, and consequently, clusters with two members will always be assigned a p-value of $p=1.0$.


\paragraph{Significance of recovered planted clusters}
We further assessed whether TiCoNE's p-values could be used to discriminate between planted enriched clusters and random clusters. We therefor generated data sets of types V1-V4, and clustered them into $k=10$ and $k=30$ clusters, while manually forcing enriched clusters into the clustering. We then used permutation functions $R1-R5$ appropriate for the respective data types, to derive cluster p-values.

As can be seen in \autoref{fig:pvalues-hidden}, all permutation functions assign small p-values to planted enriched clusters and larger, uniformly distributed p-values to background random clusters. When inspecting the false positive and false negative rates (FPR and FNR) for varying levels of $\alpha$ one can see, that all permutation functions are successful at controlling the FPR for most data sets (see \autoref{tbl:error-rates-hidden}), i.e. the error to assign a significant p-value to a random background cluster. Exception to this rule are p-values derived for data set types V3 and V4, when clustered into $k=30$ clusters and with $\alpha=0.01$. However, in these cases false positive rates are successfully controlled when using $\alpha=0.05$ instead. 

Generally, p-values of random background clusters tend to be smaller for $k=10$ than they are for $k=30$. P-values of foreground clusters are similarly small for different values of $k$, and p-value distributions of foreground clusters seem very similar for different data set types as well as similarity functions.

\begin{figure*}
\centering
\includegraphics[width=\linewidth]{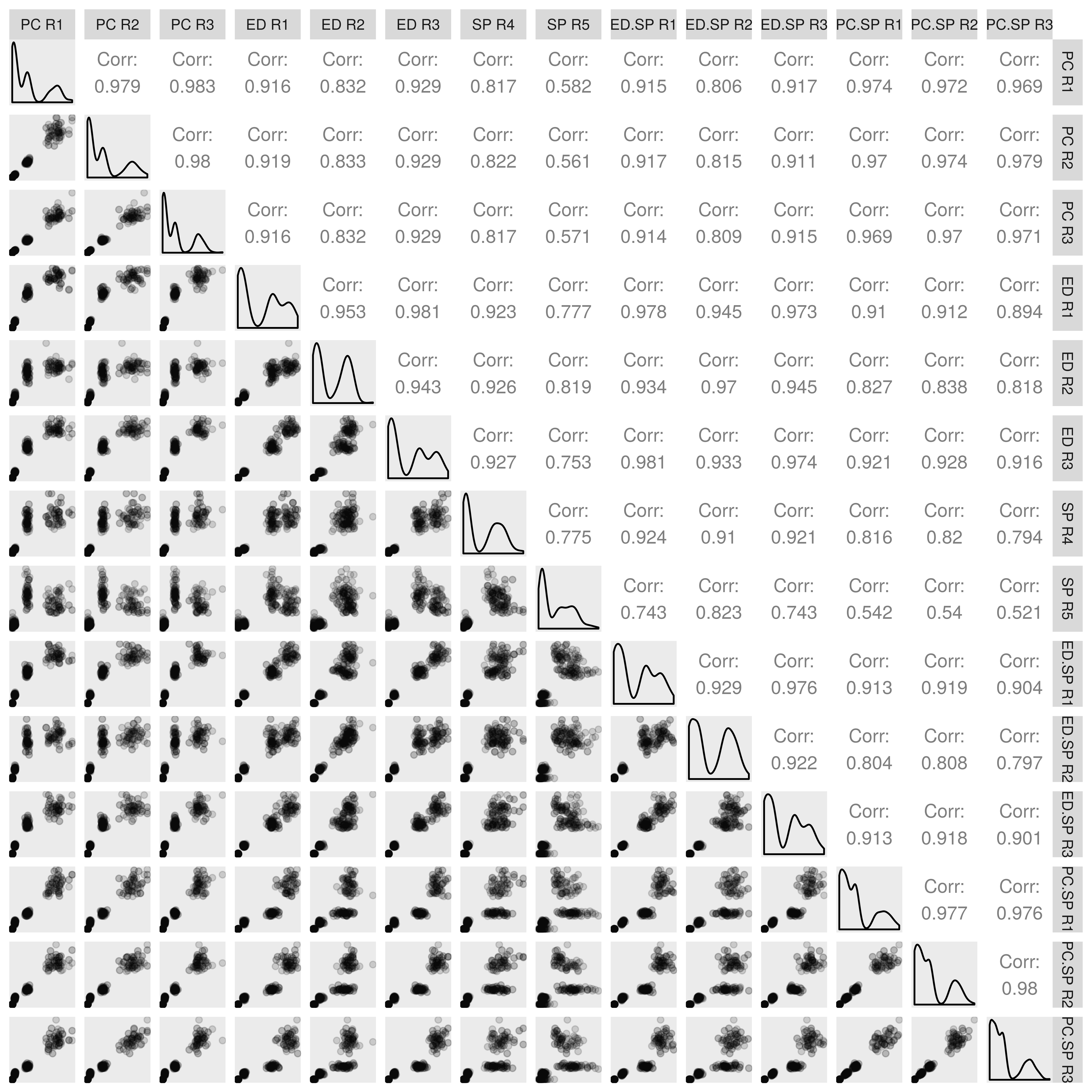}
\caption{Correlations of p-values of enriched clusters across all evaluated validation data sets. Clusters were derived using different similarity functions and p-values calculated using different permutation functions. Each cell represents p-values of two pairs $(S_1,R_1)$ and $(S_2,R_2)$ of similarity and permutation function. The upper half contains pearson correlation coefficients of p-values and the lower half contains scatter plots of the p-values of enriched clusters stemming from $(S_1,R_1)$ and $(S_2,R_2)$.}
\label{fig:pvalues-hidden-correlation-interesting}
\end{figure*}

We investigated, to what degree the p-values for the same enriched clusters derived with the different permutation functions agreed. To this end, we correlated derived p-values of these planted clusters, while excluding the random background clusters from this analysis. This is because there was no relationship between random clusters when using different similarity or permutation functions.
Correlations of p-values derived with different permutation functions but the same similarity function are, in most cases, close to $\rho\thicksim1.0$ (see \autoref{fig:pvalues-hidden-correlation-interesting}). For instance, the correlations of p-values of permutation functions R1 and R2 when using $\simed$ or $\simpc$ is $\rho=0.953$ and $\rho=0.979$ respectively. This trend also holds for composite similarity functions. For instance, the correlation of p-values derived with $\simedsp$ and R1 or R2 is $\rho=0.929$. 

Further, correlations of p-values derived with different similarity functions are close to $\rho\thicksim0.9$ for most permutation functions. This does not hold for the shortest path similarity function $\simsp$ however: It only correlates with $\rho\in[0.561,0.822]$ with $\simpc$, and $\rho\in[0.753,0.927]$ with $\simed$. Interestingly, p-values derived with $\simsp$ and permutation functions R4 or R5 correlate very poorly, which is likely due to the different types of random networks that these permutation functions generate (see \autoref{sec:permutation-functions}). In particular, permutation function R5 induces a much smaller permutation space, which may prevent generation of random clusters with properties similar to the planted ones.

\section{Discussion}
We acknowledge that our conclusions are limited by the number and types of evaluated validation data sets, and the parameter values probed, but we do believe that the design of our study is sufficiently robust allowing to carefully generalize from the here presented results.

TiCoNE is a human-guided, integrative clustering approach that allows to conjointly analyze time-series and network data. In this work we present TiCoNE 2, which comes with an extended composite model that allows to directly perform clustering of both time-series and network data using composite similarity functions. This is an improvement over TiCoNE 1, in which clustering of time-series and enrichment of clusters on a network were two separate steps. Hence, the clustering solutions derived by TiCoNE 2 are more robust towards noise in either data type.

We show that our composite model generally performs comparable to non-composite models in the absence of noise in the data. When conjointly clustering time-series and network data that both contain medium to high levels of noise, our composite model clearly outperforms the non-composite ones.

Furthermore, if one data set is noise-free or has low levels of noise, our composite model performs superior as long as the noise level of the second data set is low to medium. If, however, the noise level of the second data set is very high, non-composite models using the noise-free data are likely to perform better.

Nevertheless, the performance penalty in terms of decreased Jaccard index when using a noisy data set additionally to another high quality one is acceptably low in most cases, and is outweighed by the potentially much larger performance gains if the noise level of the second data set is acceptable. Hence, we carefully recommend using two data sets in a clustering analysis, as long as both are known to fulfill at least a medium standard in quality.

Finally, we demonstrate that TiCoNE's cluster p-values are a good indication for the meaningfulness of clusters. The here presented cluster p-values were in such high agreement with the gold standard of our validation data sets, that they could be used instead to evaluate the clusters' validities. This is especially relevant in practice, where often no such gold standard is available. From the here presented results we can conclude, that TiCoNE's permutation functions are able to distinguish between foreground and background clusters, at least if their structure is similar to the ones we evaluated here.
 
Generally, we acknowledge that the here presented results highly depend on the assumption that two independent data sets agree in a large fraction of their contained information content. Naturally, if clustering time-series and network data that contain large amounts of contradicting signals, clustering performances will be negatively affected. However, if the data sets are of sufficiently high quality, the influence of the contradicting bits of information should be outweighed by the much larger fraction of agreement between the two  data sets.

\section{Conclusions}
Accounting for technical and biological noise are two of the most important aspects of any machine learning method that processes biomedical data. Since the quality of employed data sets is often suboptimal for a multitude of reasons, noise levels should always be taken into account in method development.

One approach to address noise in data is by incorporating it into the model, such that it can easier discriminate the noise from the foreground signal. With TiCoNE 2 we here present an unsupervised learning approach that chooses another way of effectively dealing with noisy data: By conjointly analyzing multiple complementing data sets together and thereby improving the signal over noise ratio.

Our here presented results demonstrate that cluster studies can largely benefit from our approach and deliver much more accurate results by analyzing two data types together. This is especially true, if at least one of the used data sets have a medium to high level of noise.

We believe that our approach is relevant for many life scientist that perform exploratory data analysis, in that it enables analysis of multiple data sets together in a user-friendly manner and thereby deriving results of higher accuracy. TiCoNE 2 is open source software, that is available publicly as a \href{http://apps.cytoscape.org/apps/ticone}{Cytoscape App} from the Cytoscape App store.

\end{multicols}

\bibliographystyle{unsrt}
\bibliography{paper}

\end{document}